\documentclass[twocolumn,aps,prb,showpacs,floatfix,epsfig]{revtex4}

\usepackage{graphicx}%
\usepackage{dcolumn}
\usepackage{bm}%

\newcommand \be{\begin{equation}}
\newcommand \ee{\end{equation}}
\newcommand \ba{\begin{eqnarray}}
\newcommand \ea{\end{eqnarray}}
\begin{document}
\title
{\bf Dynamic avalanche breakdown of a $p\,$-$n$ junction:
deterministic triggering of a plane streamer front}
\author{Pavel Rodin \cite{EMAIL} and Igor Grekhov}
\affiliation
{Ioffe Physicotechnical Institute, Politechnicheskaya 26,
191041, St.-Petersburg, Russia}

\setcounter{page}{1}
\date{\today}

\hyphenation{cha-rac-te-ris-tics}
\hyphenation{se-mi-con-duc-tor}
\hyphenation{fluc-tua-tion}
\hyphenation{fi-la-men-ta-tion}
\hyphenation{self--con-sis-tent}
\hyphenation{cor-res-pon-ding}
\hyphenation{con-duc-ti-vi-ti-tes}


\begin{abstract}
We discuss the dynamic impact ionization breakdown of high voltage
$p\,$-$n$ junction which occurs when the electric field is increased
above the threshold of avalanche impact ionization on a time scale
smaller than the inverse thermogeneration rate. The
avalanche-to-streamer transition characterized  by generation of
dense electron-hole plasma capable to screen the applied external
electric field occurs in such regimes. We argue that the
experimentally observed deterministic triggering of the plane
streamer front at the electric field strength  above the threshold
of avalanche impact ionization but yet below the threshold of
band-to-band tunneling is generally caused by field-enhanced
ionization of deep-level centers. We suggest that the
process-induced sulfur centers and native defects such as EL2,
HB2, HB5 centers initiate the front in Si and GaAs structures,
respectively. In deep-level free structures the plane streamer front
is triggered by Zener band-to-band tunneling.
\end{abstract}



\maketitle
Impact ionization generally leads to avalanche breakdown of
semiconductor $p\,$-$n$ junctions at high reverse biases. For common
breakdown mode the space charge of avalanche carriers drastically
modifies the electric field profile, but the voltage $U$ at the
$p\,$-$n$ junction remains close to the stationary breakdown voltage
$U_b$.\cite{Sze} Different breakdown mode occurs when the electric
field is rapidly increased on a time scale smaller than the
inverse rate of thermogeneration.\cite{Si,GaAs} In absence of
thermal carriers in depleted $p\,$-$n$ junction the voltage at the
structure $U$ continues to increase far above $U_b$ before impact
ionization sets in.\cite{Si,GaAs} Then $U$ rapidly decreases to a
negligible residual value $U_{\rm res} \ll U_{\rm b}$ because the
impact ionization front passes across the structure and fills it
with dense electron-hole plasma. Initiation of such front
manifests the avalanche-to-streamer transition which is known in
gas discharge \cite{RAIZER} and semiconductor physics \cite{DYA}
mostly with respect to finger-like streamers. Streamer breakdown
is characterized by full screening of the applied electric field
by the avalanche carriers and occurs when the Maxwell relaxation
time in the avalanche becomes smaller than the inverse impact
ionization rate.\cite{DYA} In high voltage  $p^{+}$-$n$-$n$ structures
the plane TRAPATT-like ionization front moves from the $p^{+}$-$n$
junction into depleted $n$ base faster than the saturated carrier
velocity $v_s$ \cite{Si,GaAs} and is generally regarded as a plane
streamer front. \cite{Ute} This process has found important
applications in pulse power electronics.\cite{applications}

In Si and GaAs structures the onset of plane streamer front occurs
at the electric field strength $F_{\rm th}$  above the threshold
of avalanche impact ionization $F_{\rm a}$ but below the threshold
of Zener band-to-band tunneling $F_{Z}$ (e.g., in Si $F_{\rm a}
\sim 2 \cdot 10^5$~V/cm and  $F_{\rm Z} \sim 10^6$~V/cm, whereas
$F_{\rm th}\sim 3 \cdot 10^5$~V/cm).\cite{Si,GaAs} Remarkably, the breakdown sets
in a pronounced deterministic way: triggering voltage and time at
identical driving circuit conditions are not only  reproducible,
but their variation (jitter) is even below the measurement
accuracy.\cite{Si,GaAs} This implies the existence of a certain
deterministic field-dependent mechanism which supplies the initial
carriers to the depleted part of the structure. This mechanism is
not yet identified (e.g., see Ref.\ \onlinecite{ROD02a} and
discussion therein). In this Letter we argue that plane streamer
fronts are initiated by the field enhanced ionization of
deep-level midgap electron traps in the depleted part of the $p\,$-$n$
junction. We suggest that in Si structure this occurs due to the
process-induced deep-level sulfur center, whereas in GaAs
structures the native defects such as EL2, HB2 and HB5 centers are
responsible for that. In defect-free structures the plane streamer
front is initiated by band-to-band tunneling in accordance to the
recent theoretical prediction and experimental observation of
tunneling-assisted impact ionization fronts.\cite{ROD02,MESYATZ}

\begin{figure*}
\begin{center}
\hskip -0.5cm
\includegraphics[width=5.0 cm,height=7.0 cm,angle=270]{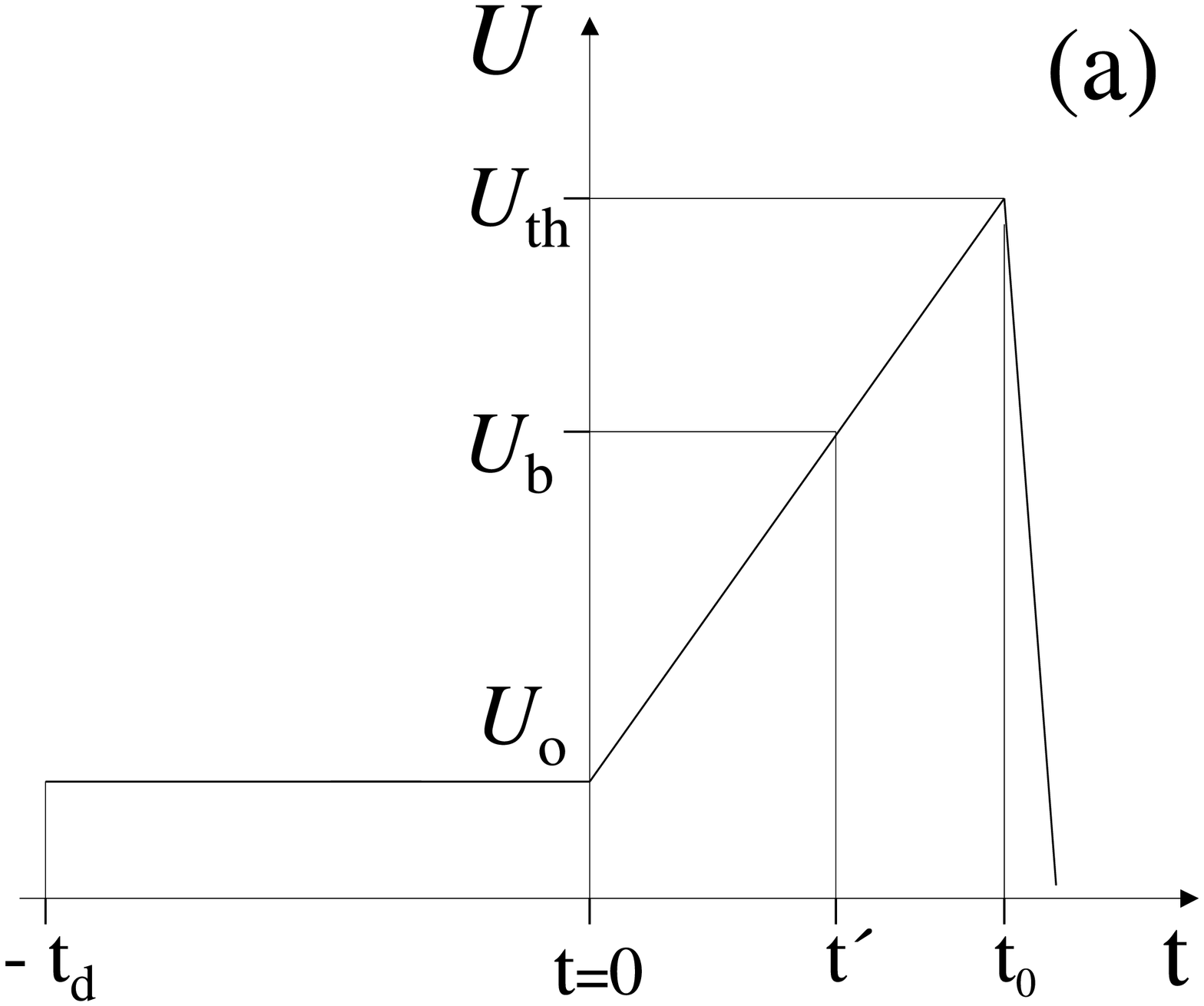}
\hskip 2.5cm
\includegraphics[width=5.0 cm,height=7.0 cm,angle=270]{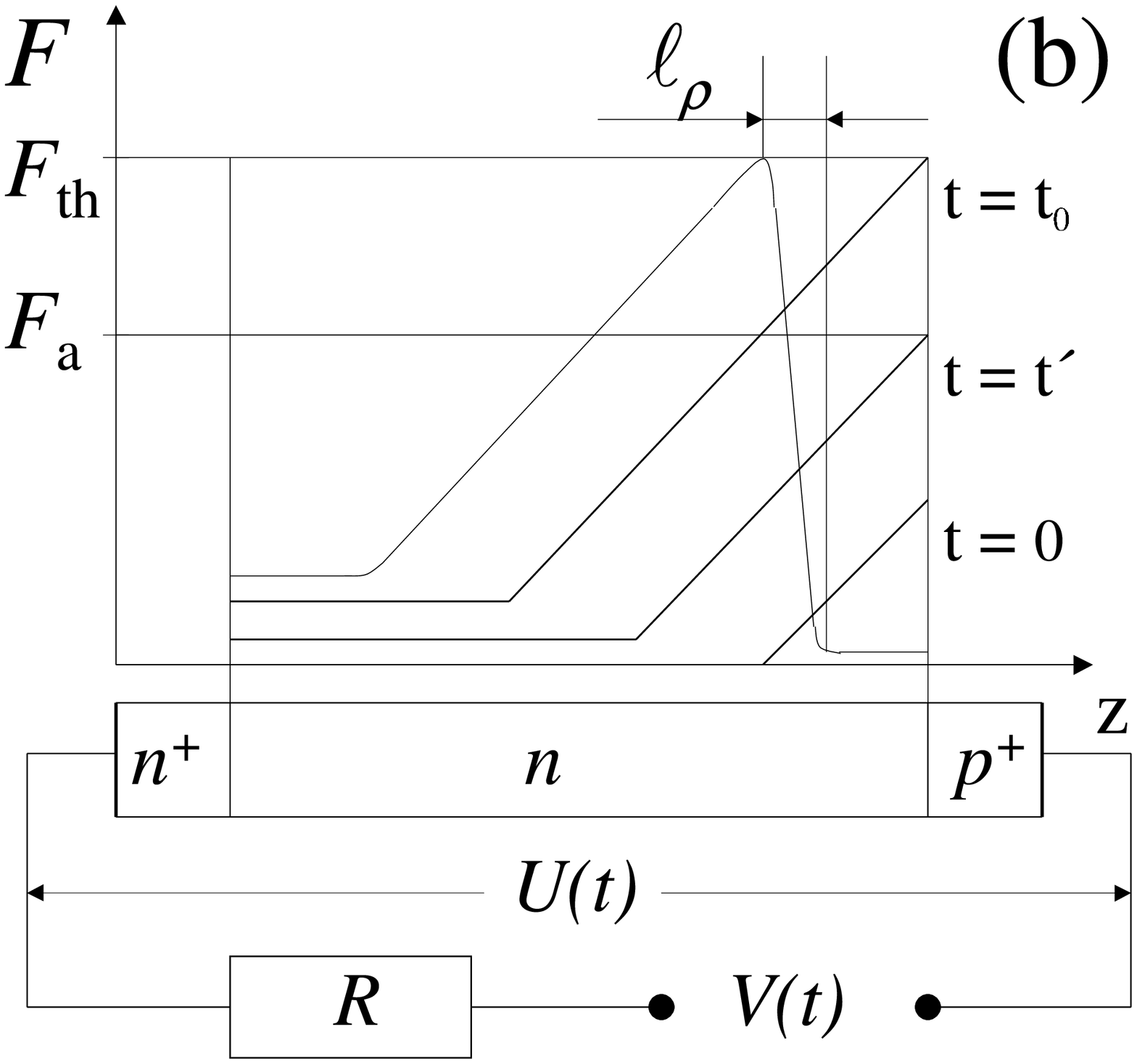}
\end{center}
\caption {Sketches of the voltage at the $p^{+}$-$n$-$n^{+}$ structure
$U(t)$ (a) and distributions of the electric field $F(z,t)$ in the
$n-$base (b). For simplicity  we assume that $V(t) \equiv U(t)$
during the period  $0 < t < t_0$, neglecting the displacement
current. The electric field profile is shown at $t=0$, at
$t=t^{\prime}$ when the voltage of stationary breakdown $U_b$ is
reached and at $t=t_0$ when the front is triggered. $F_{\rm a}$
and $F_{\rm th}$ are the effective threshold of avalanche impact
ionization and the maximum electric field achieved at the moment
when the front is triggered. The thin line in panel (b) shows the
field profile in the traveling front for $t>t_0$. The width of the
charged layer at the front tip that screens the electric field is
denoted as $\ell_{\rho}$.}
\label{sketch_voltage}
\end{figure*}

Let us consider a typical high voltage $p^{+}$-$n$-$n^{+}$ structure
with low-doped $n$ base ($N_d \sim 10^{14} \; {\rm cm^{-3}}$ for
Si and $10^{16} \; {\rm cm^{-3}}$ for GaAs) of a characteristic
width 100~${\rm\mu m}$. \cite{Si,GaAs} At the very beginning the
structure is in equilibrium. The midgap electron levels in the $n$
base are expected to be occupied. Then the initial reverse bias
$U_0$ is quasistatically applied to the structure connected in
series with load resistance $R$. During the waiting period $-t_d <
t < 0$ [Fig.\ 1(a)] some deep centers may emit electrons, but the
emission rate $e$ at low electric fields is too small to change
the occupation of midgap centers significantly. At $t=0$ the steep
voltage ramp $V(t)=U_0+A \cdot t$ is applied, where $A$ is
typically of the order of 1...10~kV/ns.
\cite{Si,GaAs,ROD02,MESYATZ} The electric field $F$ in the $n$
base increases and overcomes the effective threshold of avalanche
impact ionization $F_{\rm a}$ on a nanosecond scale [Fig.\ 1(b)].
However due to the absence of initial carriers the impact
ionization does not occur immediately and the electric field keeps
increasing. The ionization probability of deep levels $e(F)$ is
stimulated by electric field due to the phonon-assisted tunneling
mechanism that proceeds to direct tunneling mechanism in higher
electric fields.\cite{ABA91} Since $e(F)$ dependence is very
steep, the emission of free carriers abruptly starts at a certain
electrical field $F > F_{\rm a}$ triggering propagation of a plane
streamer front at $t=t_0$. To ensure deterministic triggering, the
initial carriers should be distributed uniformly over the device
area on a scale determined by an internal characteristic dimension
of the traveling front. This scale can be defined as a width
$\ell_{\rho} \sim 10 \; {\rm \mu m}$ (Ref.\ \onlinecite{ROD02a})
of the charged layer between the high-field nonconductive region
and the low-field conductive region [Fig.\ 1(b)]. Hence $n_0 \sim
\ell_{\rho}^{-3} \sim 10^{9} \; {\rm cm^{-3}}$ estimates the
threshold concentration of initial carriers.

Two decades after the effect had been discovered, it was realized
that the common Soviet fabrication technology \cite{GRE64} used to
manufacture high voltage Si structures exhibiting superfast
switching \cite{Si} creates a certain deep-level defect in the
low-doped $n$ base.\cite{sulfur} This process-induced (PI) defect,
previously known in Si and regarded as a recombination
center,\cite{Sah} later on was proved to be an electron trap with
negligible  recombination activity, thus having no effect on the
carriers lifetime, and identified as sulfur (S) impurity
center.\cite{sulfur} This double donor (see inset in Fig.\ 2) with
ionization energies $(E_c-0.28)$~eV for the upper midgap $S^0$
state (U level) and $(E_c-0.54)$~eV for the midgap $S^{+}$ state
(M level) appears in concentrations $N_{PI} = 10^{11}...10^{13} \;
{\rm cm^{-3}}$.\cite{sulfur} Taking into consideration the
field-enhanced ionization of PI centers, we are able to resolve
the  long-standing puzzle of deterministic triggering of plane
streamer fronts in Si structures. The emission rate $e(F)$ of this
PI center has been experimentally measured only for $F$ up to
$10^5$~V/cm,\cite{Sah} so we use the results of the semiclassical
theory \cite{ABA91} to evaluate $e(F)$ for the relevant field
strength $F \sim 2...4 \cdot 10^5$~V/cm. At room temperature  only
M levels (0.54~eV) are  occupied. The curve 1 in Fig.\ 2 presents
the emission rate $e(F)$ for the M level at $T = 300$~K calculated
according to the  analytical expression \cite{ABA91}
\begin{eqnarray}
e(F)=e(0) \exp \left(\frac{F^2}{F_c^2}\right)
\exp \left(\frac{2 \sqrt{2 m E_{\rm B}}}{q \tau_2 F} \ln \frac{12 F^2}{F_c^2}
\right), \\
\nonumber
F_c^2 \equiv {3 m \hbar}/({q^2 \tau_2^3}), \; \;
{2 \tau_2}/{\hbar} \equiv {1}/{kT} + {2 \tau_1}/{\hbar} \qquad
\label{Coulomb}
\end{eqnarray}
which describes phonon-assisted tunneling with the positive charge
of the ionized double donor taken into account.  Here $e(0)
\approx 2 \cdot 10^2 \; {\rm s^{-1}}$ is the emission rate in zero
electric field, \cite{sulfur,Sah} $m$ and $q$ are the effective
mass and the charge of the electron, $k$ is Boltzmann constant,
$T$ is the lattice temperature, $E_{\rm B}$ is the Bohr energy,
$\tau_1$ is the tunneling time of the vibrational subsystem
determined by the energy $\varepsilon_{\rm ph}$ of the local
phonon mode $\varepsilon_{\rm ph}/k \sim \hbar/ 2 k \tau_1 \sim
1000$~K.\cite{ABA91} We find that $e(F) \sim 10^6 \; {\rm s^{-1}}$
for $F \sim 3 \cdot 10^5$~V/cm (Fig.\ 2). For $N_{PI} = 10^{12} \;
{\rm cm^{-3}}$ the total emission rate is  $G = N_{PI} e(F) \sim
10^{18} \; {\rm s^{-1} cm^{-3}}$, and the initial concentration
$n_0 \approx 10^{-9} \; {\rm cm^{-3}}$ can be reached within
$\Delta t \approx n_0 / N_{PI} e(F) \sim 1$~ns. This is about the
time it takes for the voltage $U$ to increase from the stationary
breakdown voltage $U_{\rm b} \approx 2$~kV to the triggering
voltage $U_{\rm th} \approx 3$~kV for the typical voltage ramp $A
= 1$~kV/ns.\cite{Si} Hence for the room temperature the front can
be deterministically triggered due to the phonon-assisted
tunneling of electrons bound on the M level of the PI center. The
emission rate $e(F)$ increases with $T$, and for $T \gtrsim 350$~K
bound electrons are prematurely emitted in electrical fields below
$F_a$. This prevents triggering the streamer front and leads to
common avalanche breakdown.

For low temperatures $T \lesssim 200$~K all PI centers are
in the ground state: the U level is occupied and the M level is
empty. Curve 2 in Fig.\ 2 shows the emission rate for the U level
at low temperature when ionization occurs via direct tunneling.
The rate is evaluated according to \cite{ABA91}
\begin{eqnarray}
e(F)= \frac{q F}{\sqrt{8 m E_0}} \exp \left(-\frac{F_{0}}{F} \right)
\exp \left(2 \sqrt{\frac{E_B}{E_0}} \ln \frac{6 F_{0}}{F} \right),
\label{direct}
\\ \nonumber
F_0 \equiv {4 \sqrt{2 m E_0^3}}/{3 q \hbar}. 
\qquad \qquad \qquad \qquad \qquad \qquad \; \;
\end{eqnarray}
\begin{figure}
\begin{center}
\includegraphics[width=5.5 cm,height=7.0 cm,angle=270]{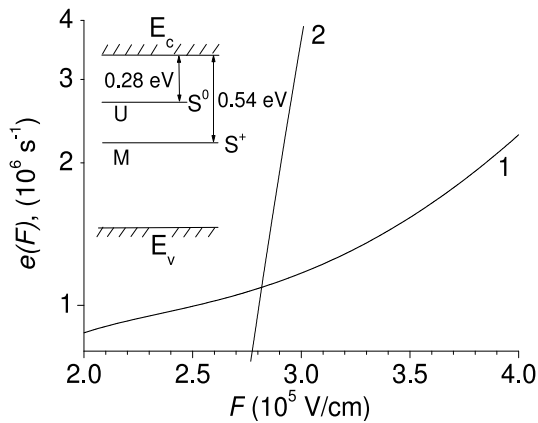}
\end{center}
\caption {The ionization rate $e$ of the midgap M level (0.54~eV)
at T=300~K (curve 1) and upper midgap U level (0.28~eV) at T=77~K
(curve 2) of the process-induced deep-level sulfur center in Si as
a function of the electric field strength $F$. The inset
illustrates the position of respective levels in the gap and the
charge states of the double donor S impurity.}
\label{Si}
\end{figure}
Here $E_0=0.28$~eV is the energy of the respective deep level.
Similar to the room temperature case, we find $e(F) \sim 10^6 \;
{\rm s^{-1}}$ at $F \sim 3 \cdot 10^5$~V/cm, although the $e(F)$
dependence is much more steep. We conclude that at low
temperatures the front can be deterministically triggered due to
direct tunneling of electrons bound on the U level. This
triggering mechanism remains unchanged up to zero temperature. In
experiments deterministic triggering of superfast fronts has been
shown to disappear for $T > 370$~K and proven to sustain at low
temperatures up to 77~K, \cite{GRE81} in agreement with our
analysis. Details of the calculations will be reported elsewhere.
Our findings suggest that concentration of deep-level sulfur
centers should be carefully controlled in fabrication of Si power
devices used in pulse sharpening applications \cite{applications}
to ensure efficient and reliable operation.

Unlike high purity Si, where the existence of deep-levels that are
capable to trigger the superfast front should be regarded as a
rare lucky exception, semi insulating GaAs is a compensated
material which possesses deep levels in large concentrations.
\cite{EL2} The most known defect is the EL2 deep-level donor
($E_c-0.75$~eV) which is present in concentration $\sim 10^{16} \;
{\rm cm^{-3}}$ for free carrier concentration up to $10^{15} \;
{\rm cm^{-3}}$ in $n$ type GaAs. \cite{EL2} In $p^{+}$-$p\,$-$n$-$n^{+}$
structures studied in Ref.\ \onlinecite{GaAs} deep-level acceptors
HL5 ($E_v+0.41$~eV) and HB2 ($E_v+0.68$~eV)\cite{EL2} are present
in concentrations $10^{14}...10^{15} \; {\rm cm^{-3}}$.
\cite{rozkov} Below we consider the EL2 defect to illustrate
possible triggering mechanism in GaAs-based $p^{+}$-$n$-$n^{+}$
structures. Due to the exact midgap position the ${\rm EL2}$ level
in $n$ type GaAs is occupied in equilibrium. It remains occupied
after any realistic waiting period $t_d$ (typically $t_d \sim 100
\; {\rm \mu s}$) since its lifetime at zero-electric field is
about 10~s.\cite{MAK82} The ionization rate $e(F)$ for the EL2
defect is known for electrical fields up to $4 \cdot 10^5$~V/cm.
\cite{MAK82,ABA91} According to Ref.\ \onlinecite{MAK82} $e(F)
\sim 10^3 \; s^{-1}$ at $F = 4 \cdot 10^5$~V/cm. For $N_{EL2} =
10^{16} \; {\rm cm^{-3}}$ the  threshold initial concentration
$n_0$ can be reached within $\Delta t \approx n_0/ N_{EL2} e(F)
\sim 100$~ps, in a reasonable agreement with
experiments.\cite{GaAs}

In deep-level-free structures the threshold of the Zener band-to-band
tunneling $F_{Z}$ can be reached, provided that the voltage ramp $A$
is sufficiently large.
In this case the plane streamer front is initiated by band-to-band tunneling
and propagates due to  the combined action of  band-to-band tunneling and impact
ionization. Such tunneling-assisted impact ionization fronts have been
described theoretically \cite{ROD02}
and  observed experimentally \cite{MESYATZ}.

Acknowledgements.-- We are indebted to E. Astrova, I. Merkulov, V.
Perel', A. Rodina and A. Rozkov for enlightening discussions. This
work was supported by the Russian Program for the Support of
Research Schools. P. Rodin is grateful to A. Alekseev for
hospitality at the University of Geneva and acknowledges the
support of the Swiss National Science Foundation.



\begin{thebibliography}{99}

\bibitem[*]{EMAIL}
Electronic address: {\rm rodin@mail.ioffe.ru}

\bibitem{Sze}
S.M. Sze, {\it Physics of Semiconductor Devices}
(Wiley, New York, 1991).

\bibitem{Si}
I.V. Grekhov and A.F. Kardo-Sysoev,
Sov.~Tech.~Phys.~Lett. {\bf 5}, 395 (1979).

\bibitem{GaAs}
Zh.I. Alferov, I.V. Grekhov, V.M. Efanov, A.F. Kardo-Sysoev,
V.I. Korol'kov, and M.N. Stepanova
Sov.~Tech.~Phys.~Lett. {\bf 13}, 454 (1987).

\bibitem{RAIZER}
E.M. Bazelyan and Yu.P. Raizer, Spark Discharges (CRS, New York, 1998)

\bibitem{DYA}
M.I. D'yakonov and V.Yu. Kachorovskii, Sov. Phys. JETP 67, 1049 (1988).

\bibitem{Ute}
U. Ebert, V. van Saarloos, and C. Caroli, Phys. Rev. E {\bf 49}, 1530 (1997).

\bibitem{applications}
I.V. Grekhov, Solid-State Electron. {\bf 32}, 923 (1989);
R.J. Focia, E. Schamiloghu, C.B. Fledermann, F.J. Agee
and J. Gaudet, IEEE Trans.~Plasma~Sci. {\bf 25}, 138 (1997).

\bibitem{ROD02a}
P. Rodin, U. Ebert, W. Hundsdorfer and I. Grekhov,
J.~Appl.~Phys. {\bf 92}, 1971 (2002).

\bibitem{ROD02}
P. Rodin, U. Ebert, W. Hundsdorfer and I. Grekhov,
J.~Appl.~Phys. {\bf 92}, 958 (2002).


\bibitem{MESYATZ}
S.K. Lyubutin, S.N. Rukin, B.G. Slovikovsky, S.N. Tsyranov,
Tech.~Phys.~Lett. {\bf 31}, 196 (2005).


\bibitem{ABA91}
V.N. Abakumov, V.I. Perel', and I.N. Yassievich,
{\it Nonradiative recombination in semiconductors,}
edited by V. M. Agranovich and A. A. Maradudin,
Modern Problems in Condenced Matter Sciences Vol.33,
(North-Holland, Amsterdam, 1991).

\bibitem{GRE64}
I.V. Grekhov {\it et al}, {\it The method to create a source of Al diffusion
in Si}, USSR patent N 176989, priority from July 6th 1964.

\bibitem{sulfur}
E.V. Astrova, V.B. Voronkov, V.A. Kozlov and A.A. Lebedev,
Semicond.~Sci.~Technol. {\bf 13}, 488-495 (1998).

\bibitem{Sah}
L.D. Yau and C.T. Sah, Solid-State Electronics {\bf 17},
193 (1974);
J.~Appl.~Phys. {\bf 46}, 1767 (1975).

\bibitem{GRE81}
I.V. Grekhov, A.F. Kardo-Sysoev, L.S. Kostina,
and S.V. Shenderey,
Sov.~Phys.~Tech.~Phys. {\bf 26}, 984 (1981).

\bibitem{EL2}
J.C. Bourgoin, H.J. von Bardeleben, and D. Stie'venard,
J.~Appl.~Phys. {\bf 64}, R65 (1988).

\bibitem{MAK82}
S. Makram-Ebeid, M. Lannoo,
Phys. Rev. {\bf B 25}, 6406 (1982).

\bibitem{rozkov}
V.I. Korol'kov, A.V. Rozkov, F.Yu. Soldatenkov and K.V. Yevstigneyev,
{\it Investigation of temperature switching stability of
AlGaAs/GaAs-based high voltage superfast switches,}
Proceedings of the 4th International Seminar on Power Semiconductors
ISPS'98, Prague, 2-4 September 1998, pp.163-168.


\end{thebibliography}
\end{document}